\documentclass[twocolumn,tighten,times]{aastex63}
\usepackage[utf8]{inputenc}
\usepackage{natbib}
\usepackage{times}
\usepackage{MnSymbol}
\usepackage{multirow}
\usepackage{amssymb,amsmath}
\usepackage[caption=false]{subfig}

\newcommand{\kps}{{\rm km}\,{\rm s}^{-1}}
\newcommand{\rgc}{$R_{\rm M49}$}

\newcommand{\reff}{$r_{\rm eff}$}
\newcommand{\vrp}{$<\!v_{r,los}\!>$}
\newcommand{\vdrp}{$<\!\sigma_{r,los}\!>$}

\submitjournal{ApJ}
\accepted{\today}

\shorttitle{M49 Globular Cluster System Kinematics}
\shortauthors{Taylor et al.}

\begin{document}

\title{Fresh Insights on the Kinematics of M49's Globular Cluster System with MMT/Hectospec Spectroscopy}
\correspondingauthor{Matthew A. Taylor}
\email{matthew.taylor@nrc-cnrc.gc.ca}

\author[0000-0003-3009-4928]{Matthew A. Taylor}
\altaffiliation{National Research Council of Canada Postdoctoral Fellow}
\affil{National Research Council of Canada, Herzberg Astronomy and Astrophysics Research Centre, 5701 West Saanich Road, Victoria, BC V9E 2E7 Canada}
\author[0000-0001-6333-599X]{Youkyung Ko}
\affiliation{Korea Astronomy and Space Science Institute, 776 Daedeok-daero, Yuseong-Gu, Daejeon 34055, Korea}
\author[0000-0003-1184-8114]{Patrick C\^ot\'e}
\affil{National Research Council of Canada, Herzberg Astronomy and Astrophysics Research Centre, 5701 West Saanich Road, Victoria, BC V9E 2E7 Canada}
\author[0000-0002-8224-1128]{Laura Ferrarese}
\affil{National Research Council of Canada, Herzberg Astronomy and Astrophysics Research Centre, 5701 West Saanich Road, Victoria, BC V9E 2E7 Canada}
\author[0000-0002-2073-2781]{Eric W.\ Peng}
\affil{Kavli Institute for Astronomy and Astrophysics, Peking University, Beijing 100871, China}
\author[0000-0001-6047-8469]{Ann Zabludoff}
\affil{Steward Observatory, University of Arizona, 933 N. Cherry Ave., Tucson, AZ 85721, USA}
\author[0000-0002-0363-4266]{Joel Roediger}
\affil{National Research Council of Canada, Herzberg Astronomy and Astrophysics Research Centre, 5701 West Saanich Road, Victoria, BC V9E 2E7 Canada}
\author[0000-0003-4945-0056]{Rub\'en Sánchez-Janssen}
\affil{UK Astronomy Technology Centre, Royal Observatory, Blackford Hill, Edinburgh, EH9 3HJ, UK}
\author[0000-0002-7939-7607]{David Hendel}
\affiliation{Department of Astronomy and Astrophysics, University of Toronto, 50 St.\ George Street, Toronto, ON M5S 3H4 Canada}
\author[0000-0002-7924-3253]{Igor Chilingarian}
\affil{Center for Astrophysics --- Harvard and Smithsonian, 60 Garden St. MS09, Cambridge, MA 02138, USA}
\affil{Sternberg Astronomical Institute, M.V. Lomonosov Moscow State University, Universitetsky prospect 13, Moscow, 119234, Russia}
\author{Chengze Liu}
\affil{Department of Astronomy, School of Physics and Astronomy, and Shanghai Key Laboratory for Particle Physics and Cosmology, Shanghai Jiao Tong University, Shanghai 200240, China}

\author[0000-0002-1685-4284]{Chelsea Spengler}
\affil{Institute of Astrophysics, Pontificia Universidad Cat\'olica de Chile, Av. Vicu\~na Mackenna 4860, 780436 Macul, Santiago, Chile}
\author[0000-0003-1632-2541]{Hongxin Zhang}
\affil{CAS Key Laboratory for Research in Galaxies and Cosmology, Department of Astronomy, University of Science and Technology of China, Hefei, Anhui 230026, China}
\affil{School of Astronomy and Space Science, University of Science and Technology of China, Hefei 230026, China}

\begin{abstract}
We present the first results of an MMT/Hectospec campaign to measure the kinematics of globular clusters (GCs) around M49---the brightest galaxy in the Virgo galaxy cluster, which dominates the Virgo B subcluster. The data include kinematic tracers beyond 95\,kpc ($\sim\!5.2$ effective radii) for M49 for the first time, enabling us to achieve three key insights reported here. First, beyond $\sim20\arcmin-30\arcmin$ ($\sim100-150$\,kpc), the GC kinematics sampled along the minor photometric axis of M49 become increasingly hotter, indicating a transition from GCs related to M49 to those representing the Virgo B intra-cluster medium. Second, there is an anomaly in the line-of-sight radial velocity dispersion ($\sigma_{r,los}$) profile in an annulus $\sim10-15\arcmin$ ($\sim50-90$\,kpc) from M49 in which the kinematics cool by $\Delta\sigma_{r,los}\approx150\,\kps$ relative to those in- or outward. The kinematic fingerprint of a previous accretion event is hinted at in projected phase-space, and we isolate GCs that both give rise to this feature, and are spatially co-located with two prominent stellar shells in the halo of M49. Third, we find a subsample of GCs with velocities representative of the dwarf galaxy VCC\,1249 that is currently interacting with M49. The spatial distribution of these GCs closely resembles the morphology of VCC\,1249's isophotes, indicating that several of these GCs are likely in the act of being stripped from the dwarf during its passage through M49's halo. Taken together, these results point toward the opportunity of witnessing on-going giant halo assembly in the depths of a cluster environment.
\end{abstract}


\section{Introduction} \label{sec:intro}
Low-surface brightness (LSB; $\mu_V\approx26-29\,{\rm mag}\,{\rm arcsec}^{-2}$) tidal debris like streams and shell systems are common around giant elliptical galaxies (gE's) in the nearby universe \citep[e.g.,][]{mal80}. Shell systems arise via phase mixing following the merger of one or more galaxies, typically at a high mass ratio, along radial orbits \citep[e.g.,][]{qui84}. During the merger, stars of the perturbed galaxies are jostled into new orbits by differential tidal effects and temporarily align in phase space leading to interleaved shells of stars apparent in deep imaging. While such systems can be relatively long-lived after mergers in the field, the loosely bound stellar debris can be destroyed rapidly through tidal interactions in a cluster environment \citep{mih04}.

While mergers with fainter ($M_V\ga-15$\,mag) satellites produce shell structures generally dominated by individual stars and readily traced kinematically via planetary nebulae \citep[e.g.,][]{lon15a,lon15b}, they can only be expected to contribute a handful of globular clusters (GCs) despite hosting larger numbers per unit luminosity compared to their brighter counterparts \citep[e.g.,][]{har81, geo10}. On the other hand, more massive progenitors potentially contribute richer systems of GCs to any individual event. Given enough observable GC tracers, such shells are detectable through the phase-aligned kinematics that make shells visible in the first place \citep[e.g.,][]{rom12}, but can also trace fine structure even in the absence of velocity information \citep{lim17}.

Kinematics derived from integrated diffuse stellar light is a useful tool to investigate the formation of gE's. A foundational result in this regard was provided by \citep{dre79}, who presented stellar line-of-sight radial velocity dispersions ($\sigma_{r,los}$) for the brightest galaxy in the Abell 2029 cluster, IC\,1101, out to several dozens of kpc. The average \vdrp-profile was seen to increase with radius, but fell short of the \vdrp\ of the cluster as a whole. Positive \vdrp-profile gradients have since been found to be common features of early-type galaxies in the local universe (often reaching the host cluster dispersions) and can be interpreted as sampling the transition from dynamics dominated by the host giant, to that more influenced by the intra-cluster medium (ICM) \citep[e.g.,][]{lou08, coc09, ric11, mur14, ben15, vea17, bar18}.

The challenge of studying line-of-sight velocity distributions (LOSVDs) via integrated diffuse stellar light, is that the surface-brightness of such galaxies rapidly drops beyond the core regions, so that at distant radii, spectral signal-to-noise drops to levels where the LOSVD cannot be reliably recovered. A potential solution to this is to use other kinematic tracers like GCs and PNe \citep[e.g.,][]{coc09,lon15a,lon15b,har18}, which are relatively accessible kinematic tracers that reach far out into the halos of their giant hosts, and richly populate the ICM of the clusters in which their hosts reside.

M49 is the brightest galaxy in the Virgo galaxy cluster, and dominant member of the Virgo B sub-cluster. M49 has a smooth, blue halo that was likely built up early through the accretion of low-mass \citep[${\cal M}_{\star}\la10^8\,M_\odot$;][]{har18}, low-metallicity satellites \citep[][]{mih13}. It also exhibits multiple LSB features, including a complex, interleaved shell system reaching out to 50-100\,kpc \citep{jan10, fer12}, with $(B-V)_0\approx0.85$\,mag colors that are redder than the underlying halo \citep{mih13}. Given the mean $(B-V)_0\approx0.77$\,mag color of the surrounding dwarf galaxy reservoir \citep{van04}, \cite{mih13} argue that these shells are consistent with the recent accretion of a single dwarf.

M49 is currently interacting with the dwarf galaxy VCC\,1249 \citep[e.g.,][]{arr12}, lying at a projected distance of 5.4\arcmin\ (26.2\,kpc). VCC\,1249 was imaged in $B$, $I$, and H\,I by \cite{pat92}, who found evidence for ongoing interaction, including significant ram-pressure stripping. Meanwhile, a tidal ``plume'' was found by \cite{jan10} to connect VCC\,1249 and M49, further bolstering the notion that these two galaxies are caught in the act of a high mass-ratio merger.

Here we present the first results from a wide spectroscopic survey of GCs around M49, with a focus on those likely to be associated with recent and ongoing interactions. Throughout this contribution we assume a distance to M49 of $16.7\pm0.6$\,Mpc \citep[$(m-M)_0=31.12\pm0.08$;][]{bla09}, corresponding to 4.9\,kpc per arcminute and an effective radius (\reff) of $226\pm34$\arcsec (18.3\,kpc) \citep[][]{fer20}.

%%%%%%%%%%%%%%%%%%%%%%%%%%%%%%%%%%%%%%%%%%%%%%%%%%%%%%%%%%%%%%%%%
%%%%%%%%%%%%%%%%%%%%%%%%%%%%%%%%%%%%%%%%%%%%%%%%%%%%%%%%%%%%%%%%%
%%%%%%%%%%%%%%%%%%%%%%%%%%%%%%%%%%%%%%%%%%%%%%%%%%%%%%%%%%%%%%%%%

\section{Observations and Data Processing} \label{sec:style}
\begin{figure*}
\centering
\includegraphics[width=\linewidth]{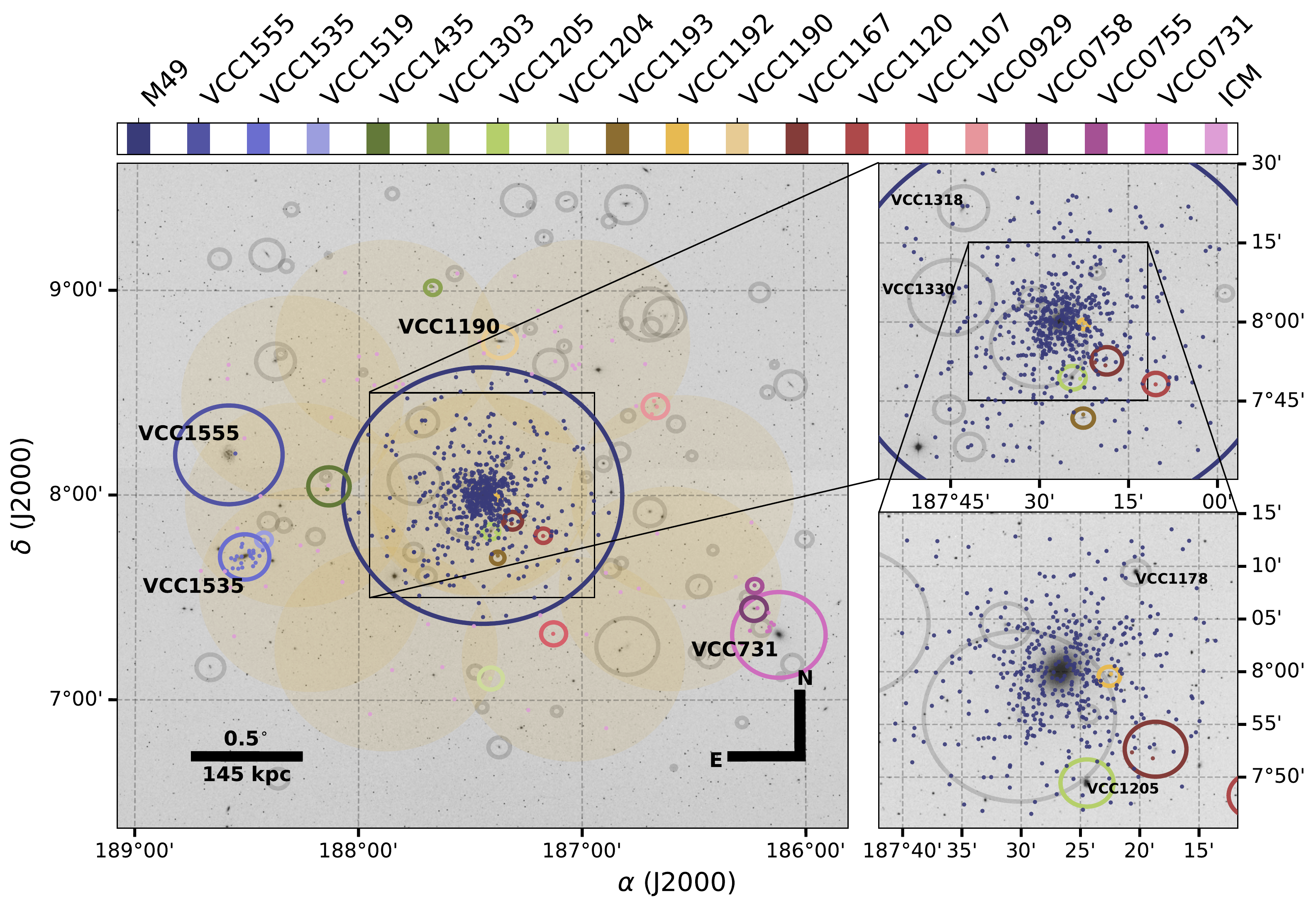}
\caption{The spatial distribution of GCs in the vicinity of M49 with measured $v_{r,los}$, overlayed on a DSS image. Dots denote GCs with colors indicating galaxy associations as determined by our classification strategy described in \S\,\ref{sec:gc_sample}, as mapped to the legend at the top of the figure. The gold shaded regions represent the ten MMT/Hectospec pointings.}
\label{fig:m49_dss}
\end{figure*}

Moderate resolution ($R=2000$) spectra were obtained between 2014 March and June, using the fiber-fed multi-object spectrometer Hectospec \citep{fab05} mounted on the 6.5\,m MMT, with 300 $1.5\arcsec$ diameter fibers available across a $1^\circ$ diameter field of view. Given the spatial extent of the intra-cluster medium (ICM) surrounding M49, we obtained spectra from ten unique pointings (Fig.\,\ref{fig:m49_dss}; orange shading). We chose a 270\,mm$^{-1}$ grating, giving a dispersion of 1.2\,\AA\,pixel$^{-1}$ across the 3650-9200\,\AA\ wavelength range. The integrated exposure time of each configuration ranges from 4500\,s to 7200\,s, covering a total of $N_{\rm obj}$ = 3015 targets.

We reduced the spectra using the HSRED pipeline\footnote{http://archive.mmto.org/node/536}. The pipeline includes bias subtraction, dark correction, flat-fielding, aperture extraction, and wavelength calibration. We adopted the cross-correlation method \citep{td79} to measure heliocentric radial velocities of the targets, using the {\it xcsao} task in the IRAF RVSAO package \citep{km98}.

%%%%%%%%%%%%%%%%%%%%%%%%%%%%%%%%%%%%%%%%%%%%%%%%%%%%%%%%%%%%%%%%%
%%%%%%%%%%%%%%%%%%%%%%%%%%%%%%%%%%%%%%%%%%%%%%%%%%%%%%%%%%%%%%%%%
%%%%%%%%%%%%%%%%%%%%%%%%%%%%%%%%%%%%%%%%%%%%%%%%%%%%%%%%%%%%%%%%%
\section{Analysis \label{sec:analysis}}
Here we describe a few aspects of our analysis, including the definition of our GC sample, and an overview of the kinematics that leads to some of the key aspects of the M49 system that we discuss in \S\,\ref{sec:results}.

%%%%%%%%%%%%%%%%%%%%%%%%%%%%%%%%%%%%%%%%%%%%%%%%%%%%%%%%%%%%%%%%%
\subsection{GC Sample Definition} \label{sec:gc_sample}
We identify GCs among the spectroscopic targets based on line-of-sight radial velocities ($v_{r,los}$) and concentration indices ($c_{ap}$). $c_{ap}$ is useful to discriminate between point-like and extended sources \citep[e.g.,][]{dur14}, and we define it as the difference between four and eight pixel aperture-corrected magnitudes measured from {\it Next Generation Virgo Survey} \citep[NGVS;][]{fer12} $g'$-band imaging. As a first cut, we consider targets showing $v_{r,los} < 3000\,\kps$ (while also meeting our $c_{ap}$ criteria) to be GCs associated with Virgo, yielding 525 GCs in the survey region. While these criteria---in particular the inclusion of morphometric information---should be very effective at filtering out foreground stellar contaminants, we take a final step to cull potential stellar sources by querying the Gaia Early Data Release 3 database \citep{gai16,gai20}. Specifically we filter those objects with proper motions ($\mu$) that imply unreasonable physical velocities at the distance of Virgo. We find 141 GCs with matched Gaia coordinates, of which 55 have $\mu$ measurements. Of these, 13 show $\mu\ga1\arcsec\,{\rm yr}^{-1}$ (translating to physical velocities $\ga10^4\,\kps$) and/or listed uncertainties inconsistent with $\mu\approx0\arcsec\,{\rm yr}^{-1}$. Despite confidence in our above selection criteria, we err on the side of caution and remove them from the subsequent analysis. We note that if all are true foreground contaminants, this suggests an upper limit of $\sim2.5$ percent contamination to our sample prior to their omission.

We desire to isolate the kinematics of M49 and the Virgo B ICM, and attempt to do so by using spatial and kinematic information to clean the sample of GCs that may be members of other galaxies in the survey footprint. Using galaxy information compiled by the NGVS, we consider a GC to belong to a particular galaxy, including M49 itself, if it is both projected within ten \reff\ of a given galaxy, and the velocity difference between a GC and the galaxy is smaller than three times the galaxy's central line-of-sight velocity dispersion \citep[$\sigma_{0,gal}$;][]{mce95}, i.e., $|v_{r,los} - v_{r,gal}| < 3\sigma_{0,gal}$. If \reff\ or $\sigma_{0,gal}$ are unknown (e.g., faint dwarf galaxies), we assume $30\arcsec$ and $50\,\kps$, respectively. Finally, if a GC satisfies the criteria to be associated with M49 {\it and} another galaxy, we assign that GC to the other galaxy (so that only those exclusively satisfying our criteria for M49 are assigned to it), and those that do not satisfy the criteria for any galaxy are assumed to belong to the ICM.

We show the results of these efforts in Fig.\,\ref{fig:m49_dss}, where circles denote $10$\reff\ of galaxies in the region, and dots indicate GCs with measured $v_{r,los}$. Colors are matched between GCs and galaxies according to the legend shown above the figure, and grey circles correspond to galaxies for which no GCs are identified. Several prominent galaxies in the region are indicated, whose GC kinematics will be analyzed in a future work. Following this procedure, we classify 58 and 415 GCs to the ICM and M49 respectively, and do not consider the remaining 52 GCs likely to be associated with other galaxies.

These observations build upon a previous spectral survey of 263 GCs sampling the innermost $\sim500\arcsec$ (\rgc$\approx40$\,kpc) around M49 \citep{cot03}. To increase sampling fidelity in this region, we incorporate their measurements in this work. We find 76 GCs to be in both catalogues, such that our total MMT sample includes 449 GCs with $v_{r,los}$ measured for the very first time. This represents a $\sim170$ percent increase over previous catalogues, sampling a ten-fold increase in M49-centric spatial extent. Focussing only on those GCs classified as either belonging to M49 or the ICM, we continue with 660 unique spectroscopic GCs.

Finally, given the $997\,\kps$ recessional velocity and $\sim300\,\kps$ dispersion of M49, there remains the potential for stellar contaminants to be falsely identified as GCs without careful vetting. This is particularly true for targets at the lower range of the GC LOSVD, and of larger concern for objects at large \rgc, where GCs tend toward bluer colors (see below) that more easily blend with the optical colors of foreground stars. On top of the initial vetting of sources using $c_{ap}$ as a discriminant, we pay extra scrutiny to those in the M49 outskirts (\rgc$\ga100$\,kpc) that also show $v_r\la100\,\kps$. We find eight GCs that satisfy these criteria and find that, after careful consideration of visible source morphology, concentration, multi-band color information, and existence (or lack) of Gaia-based proper motions, we find only two that are likely of stellar origin, and omit them from the analysis moving forward.

\begin{figure}
\centering
\includegraphics[width=\linewidth]{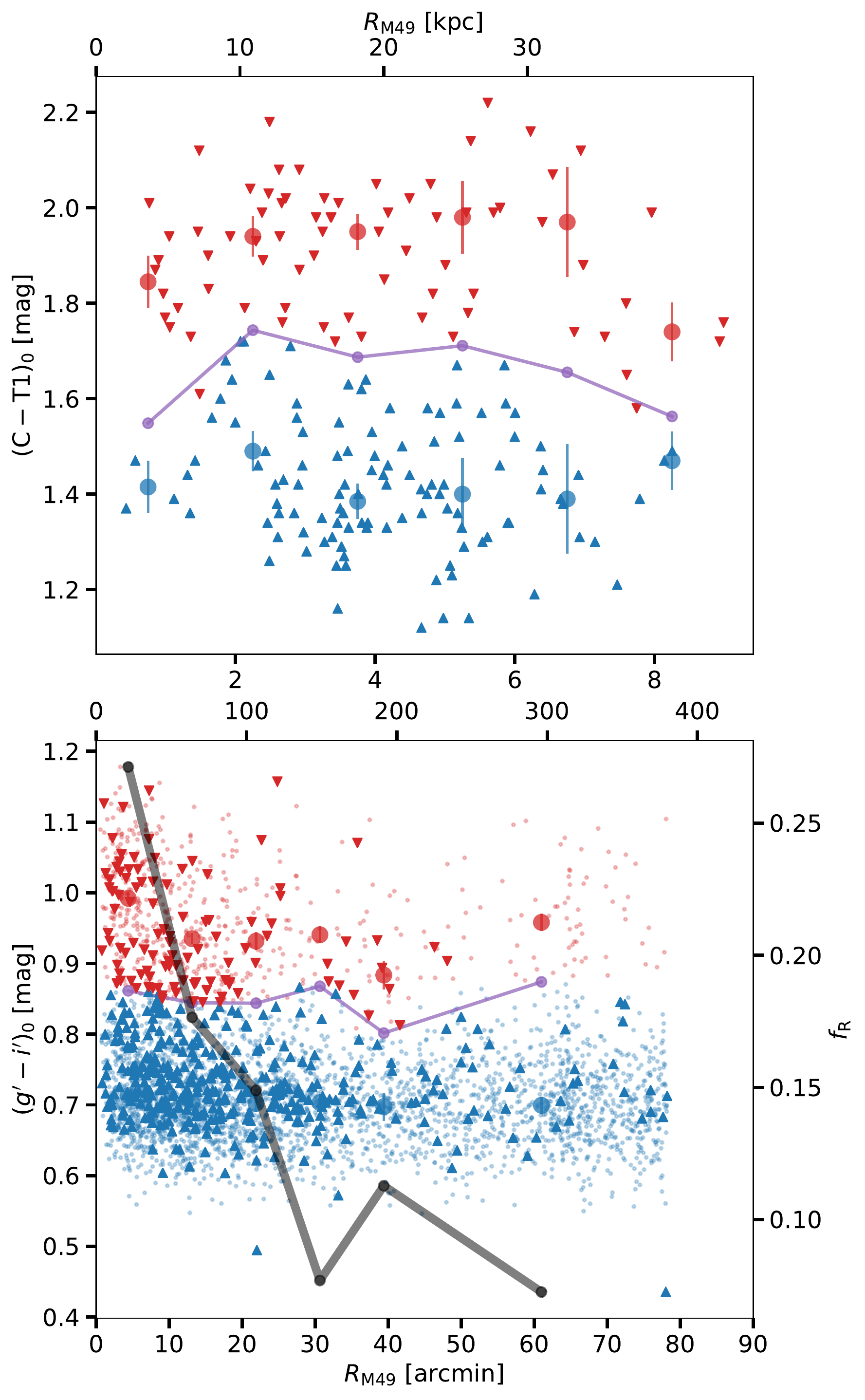}
\caption{Our GMM color selection technique fit to our photometric sample of M49 GC candidates, and applied to our spectroscopic sample. The upper panel shows the color classification fit to the sample of \cite{cot03} (triangles), using the metallicity sensitive $(C-T1)_0$ index, while below we fit to the $(g'-i')_0$ colors of 4343 GC candidates (small dots), and apply it to our spectroscopic GC sample (triangles). In each panel we show the mean red/blue colors in each radial bin by the larger colored dots, and the intersection between the two GMM components in purple. Finally, in the lower panel we show the decreasing fraction of red GCs by the black relation, corresponding to the right-hand ordinate.}
\label{fig:col_select}
\end{figure} 

\subsubsection{GC Color Classification}
We further refine the sample by classifying GCs into red (presumably metal-rich) or blue (metal-poor) subsamples. GC systems around giant galaxies show a well-known color/metallicity bimodality that reflects galactic formation histories \citep[e.g.,][]{bro06}. In one scenario, the blue/metal-poor GCs form in-situ from pristine gas alongside their host, whereas gas-rich mergers induce later starbursts that give rise to the red/metal-rich population \citep[e.g.,][]{ash92, kis97, bea02}. Alternatively, in a hierarchical assembly scenario giant galaxies develop GC systems through a two-stage process in which red/metal-rich GCs form early, deep in the potential wells and are rapidly enriched during the early, intense star formation episodes of their giant hosts. Conversely, many blue/metal-poor GCs are accreted over time through mergers with lower-mass satellites and thus generally populate the outer regions of giant halos \citep[e.g.,][]{cot98, cot00}. Either way, years of observational evidence shows that while red GCs tend to have distributions that closely reflect a giant's spheroidal light, blue GCs become more numerically dominant in the outer galaxy halos \citep[e.g.,][]{gei96, ash98, bro06, for12}.

Given the general shift toward bluer GCs in galaxy outskirts, we classify GCs by color in six radial bins as a function of projected M49-centric distance (\rgc) using a two-component Gaussian mixture model (GMM). To minimize the risk of unknown color selection biases that may have arisen during the MMT spectroscopic target selection, we incorporate 4343 GC candidates photometrically selected from the NGVS that represents the parent sample from which the spectroscopic targets were selected. In each of the six radial bins, we use the two-component GMM to classify the total sample---including the spectroscopic subsample---as either blue or red, where the classification is defined based on the intersection of the two Gaussians in each bin. Fig.\,\ref{fig:col_select} shows the results, where the distinction between GC colors is indicated by the respective symbols, and the purple relations show intersections between GMM components in each bin. The upper panel shows results for the \cite{cot03} sample where for many GCs only their metallicity-sensitive $(C-T1)_0$ color is available, and we apply the technique separately but in the same fashion as the sample shown in the lower panel.

In both panels of Fig.\,\ref{fig:col_select} the current dataset(s) are shown by triangles, while in the lower panel the larger photometric GC candidate sample is shown by fainter dots. Meanwhile, the mean colors of each bin are indicated by the larger dots with $1\sigma$ uncertainties. For either sample, there is very little variation with \rgc, with peak-to-valley color variations $\Delta(g'-i')_0\la0.1$\,mag. We find 177 red and 470 blue GCs, with the latter increasingly dominant at large \rgc. We note the rapid drop-off of red GCs in our spectroscopic sample beyond \rgc$\approx20\arcmin$ ($\approx100$\,kpc), and total lack of such beyond \rgc$\approx50\arcmin$ ($\approx240$\,kpc), the implications of which are discussed below. 

%%%%%%%%%%%%%%%%%%%%%%%%%%%%%%%%%%%%%%%%%%%%%%%%%%%%%%%%%%%%%%%%%
\subsection{One-dimensional Kinematic Profiles \label{sec:vr_profile}}
Our results are based on the first two moments (i.e., $v_{r,los}$ and $\sigma_{r.los}$) of the GC LOSVD, as profiled along the projected \rgc. To account for the ellipsoidal morphology of the M49 surface-brightness profile, we recast the projected circular-based angular distance from M49's photometric center to its ellipsoidal major axis-equivalent distance based on morphological properties measured by the NGVS, namely $r_{\rm eff,M49}=226\pm34\arcsec$ (based on a curve-of-growth analysis applied to $g'$-band NGVS imaging) and ellipticity $\epsilon=0.15\pm0.04$ with a major axis position angle at $PA=-22\pm2^\circ$ (East of North), with the latter parameters calculated from M49's isophotal center averaged between 1\arcsec\ and one $r_{\rm eff,M49}$.

\begin{figure*}
\centering
\includegraphics[width=\linewidth]{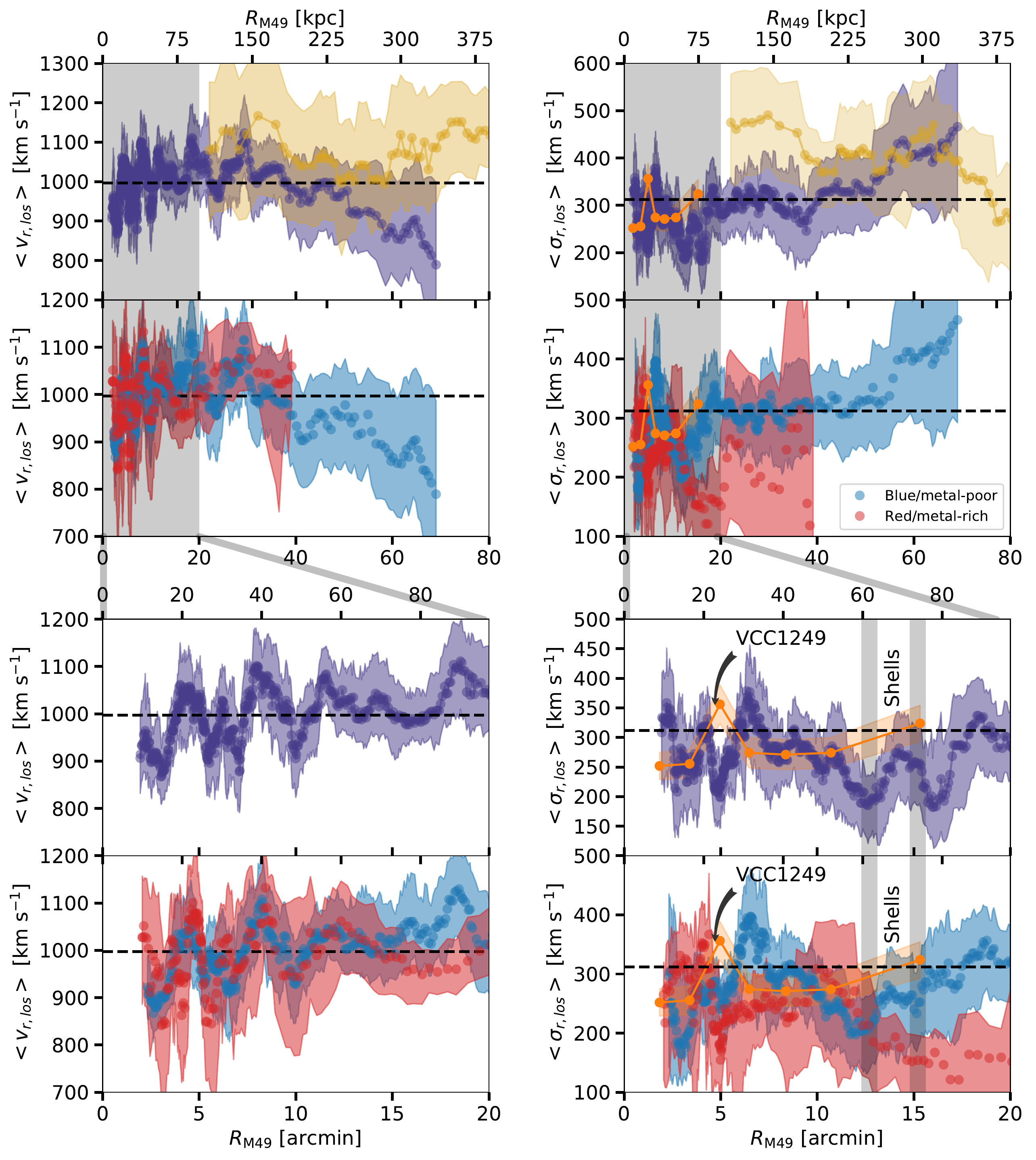}
\caption{One-dimensional radial \vrp\ (left) and \vdrp\ (right) profiles, with the bottom quartet of panels representing a zoom in of the top. Running mean kinematic profiles are shown for the M49+ICM GC sample (purple; top panels in each quartet), and red/blue GC subsamples (colored accordingly; bottom panels in each quartet). To place the GC kinematics in the context of the overall Virgo B subcluster, gold points/shading shows the running mean \vrp\ and \vdrp\-profiles for Virgo B galaxies, with data taken from the NGVS. For comparison, we point out the locations of VCC\,1249 and associated stellar shells, while the orange relation shows the \vdrp\ profile for PNe in the region from \cite{har18}. Dashed black lines indicate the M49's systemic $v_{r,los}$. Overall, we find quite complex GC kinematics interior to $\sim20\arcmin$, outside of which the profiles become smoother and show hints of merging with the Virgo B ICM.}
\label{fig:vr_profile}
\end{figure*}

Fig.\,\ref{fig:vr_profile} shows the overall one dimensional kinematic profile of M49, where the running mean \vrp- and \vdrp-profiles are shown in the left- and right-hand columns, respectively. The profiles shown in purple reflect the combined M49+ICM GC samples (see \S\,\ref{sec:gc_sample}) with no consideration for GC color, below which we show the same for the red and blue subsamples. Furthermore, we show in the lower quartet of panels a zoom-in to the innermost $20\arcmin$ to highlight the dynamics in the immediate vicinity of M49. For each profile, we radially bin the data centered on each GC in turn, such that each bin contains the 35 nearest (in terms of 1D \rgc) GCs---15 for the red sub-sample, on account of their lower numbers---from which the bi-weight mean \vrp\ and \vdrp\ are calculated and plotted on Fig.\,\ref{fig:vr_profile} as a function of average \rgc. Similarly, and to place these data in the context of the overall Virgo B subcluster, we show by the gold points/shading the running mean profiles of galaxies strewn through the sampled region. In all cases, shading that brackets the data represents the $2\sigma$ uncertainties calculated via 1000 bootstrap realizations.

%%%%%%%%%%%%%%%%%%%%%%%%%%%%%%%%%%%%%%%%%%%%%%%%%%%%%%%%%%%%%%%%%
%%%%%%%%%%%%%%%%%%%%%%%%%%%%%%%%%%%%%%%%%%%%%%%%%%%%%%%%%%%%%%%%%
%%%%%%%%%%%%%%%%%%%%%%%%%%%%%%%%%%%%%%%%%%%%%%%%%%%%%%%%%%%%%%%%%

\section{Results \& Discussion} \label{sec:results}

A complete analysis and discussion of the GC kinematics across the entire field of view probed by the MMT/Hectospec data is reserved for a future work, but here we comment on three noteworthy features in Fig.\,\ref{fig:vr_profile} that are described and further explored below.

\subsection{Transition to the Intra-cluster Medium}
We first note the marked decline in the fraction of red GCs ($f_{R}$) out to \rgc$\approx20-30\arcmin$ ($\approx100-150$\,kpc) that can be seen in Figs.\,\ref{fig:col_select} and \ref{fig:vr_profile}, which is illustrated by the black relation in Fig.\,\ref{fig:col_select} (lower panel). Such a feature is observed to be common in the GC systems of giant galaxies \citep[e.g.,][]{gei96, ash98, bro06, for12}, whether built up via the merger \citep[e.g.,][]{ash92, kis97, bea02} or hierarchical \citep[e.g.,][]{cot98, cot00} scenarios. In the two-stage hierarchical model of galaxy mass assembly a radial decline in $f_R$ is to be expected, as these GCs are expected to form early, deep in M49's potential well. Likewise, the transition to bluer GCs is consistent with later accretion of GCs from interactions with low-mass satellites and/or those thrown into the ICM by the dynamically violent environment. We note that this latter interpretation is not conclusive, as in at least some giant GC systems, the metallicities of metal-poor GCs differs from that of GCs associated with dwarf galaxies \citep[e.g.,][]{str04}, and we continue with this point in mind.

\begin{figure}
\centering
\includegraphics[width=\linewidth]{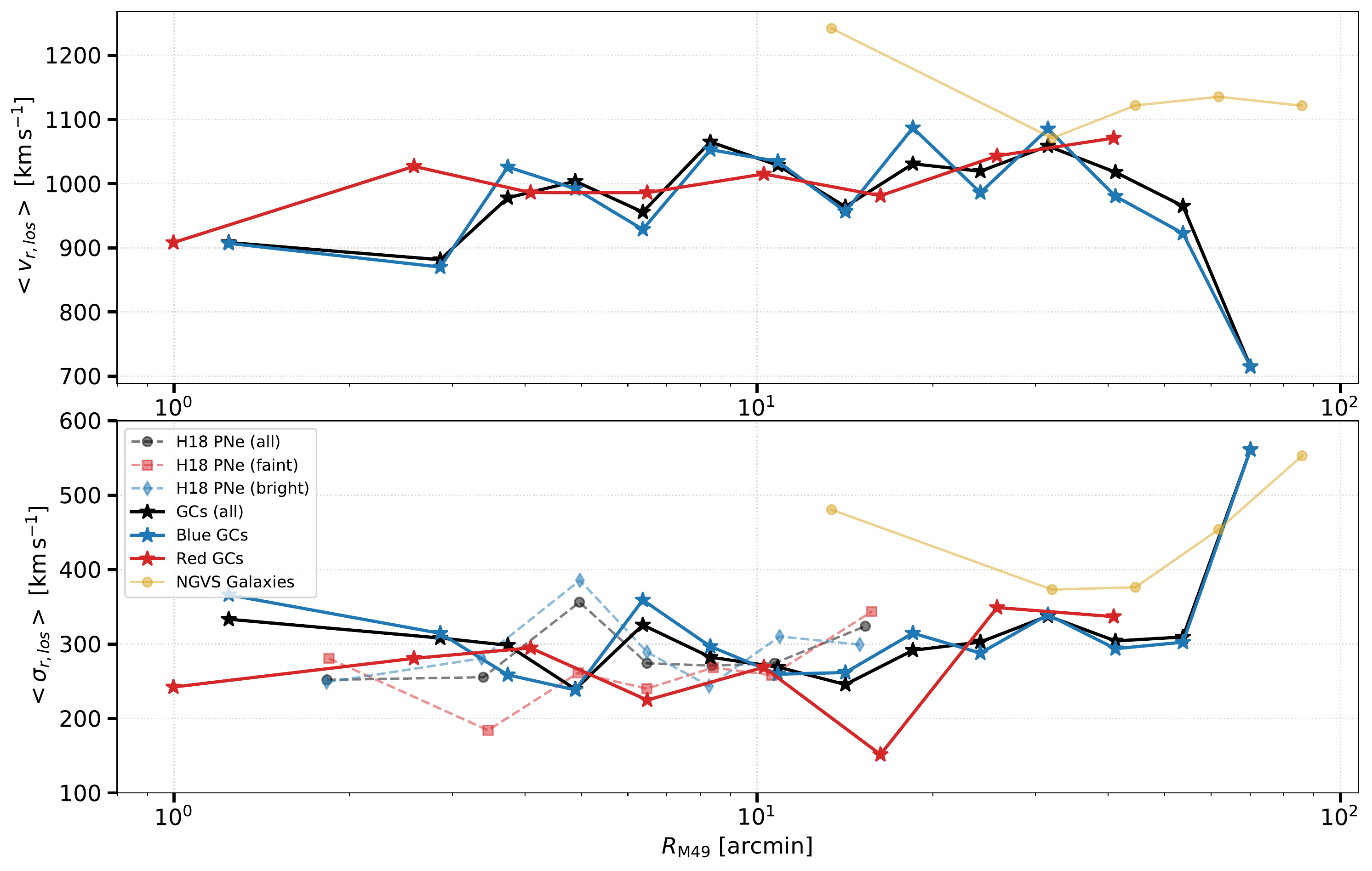}
\caption{The \vdrp\ profile computed in radial bins for direct comparison to the PNe kinematics from \cite{har18}. The top panel shows the \vrp\ profile for the present GC sample (solid blue, red, and black relations), while the bottom shows \vdrp\ with the dashed lines indicating the profile from \cite{har18}, split into their subsamples following the figure legend. In both panels we show the binned profiles of Virgo B subcluster galaxies by the gold relations.}
\label{fig:h18_comp}
\end{figure} 

We are interested in whether the kinematic profiles in this region hint at a transition from GCs intrinsic to M49 to those dominated by the dynamics of the ICM of the Virgo B subcluster. Fig.\,\ref{fig:h18_comp} shows a direct comparison to the results of \cite{har18}, who recently reported the transition to the ICM through a study of M49 planetary nebulae (PNe), finding that their dynamics begin to merge with that of the ICM near \rgc$\approx100$\,kpc. This result is consistent with large-scale positive \vdrp\ gradients seen in the PNe kinematics of M49 \citep[][]{har18, pul18}, but also in the outer halos of other giant galaxies such that the tracer kinematics merge with that of the surrounding ICM and indicates the increasing influence of the local ICM \citep[e.g.,][]{dre79,ben15, bar18}. We find a similar signature in the transition from the complex dynamics at \rgc$\la100$\,kpc, to the smoother \vdrp\ profile beyond, where Fig.\,\ref{fig:vr_profile} shows the dispersion to increase almost monotonically with \rgc\ from a relatively flat \vdrp$\approx300\,\kps$ to the \vdrp$\approx400-500\,\kps$ level more consistent with the gold points/shading that indicate the \vdrp-profile arising from Virgo B subcluster giants. Based on this, these results bolster those of \cite{har18}, in that beyond \rgc$\,\approx\!100$\,kpc, GC dynamics become increasingly representative of the Virgo B subcluster ICM.

\begin{figure*}
\centering
\includegraphics[width=\linewidth]{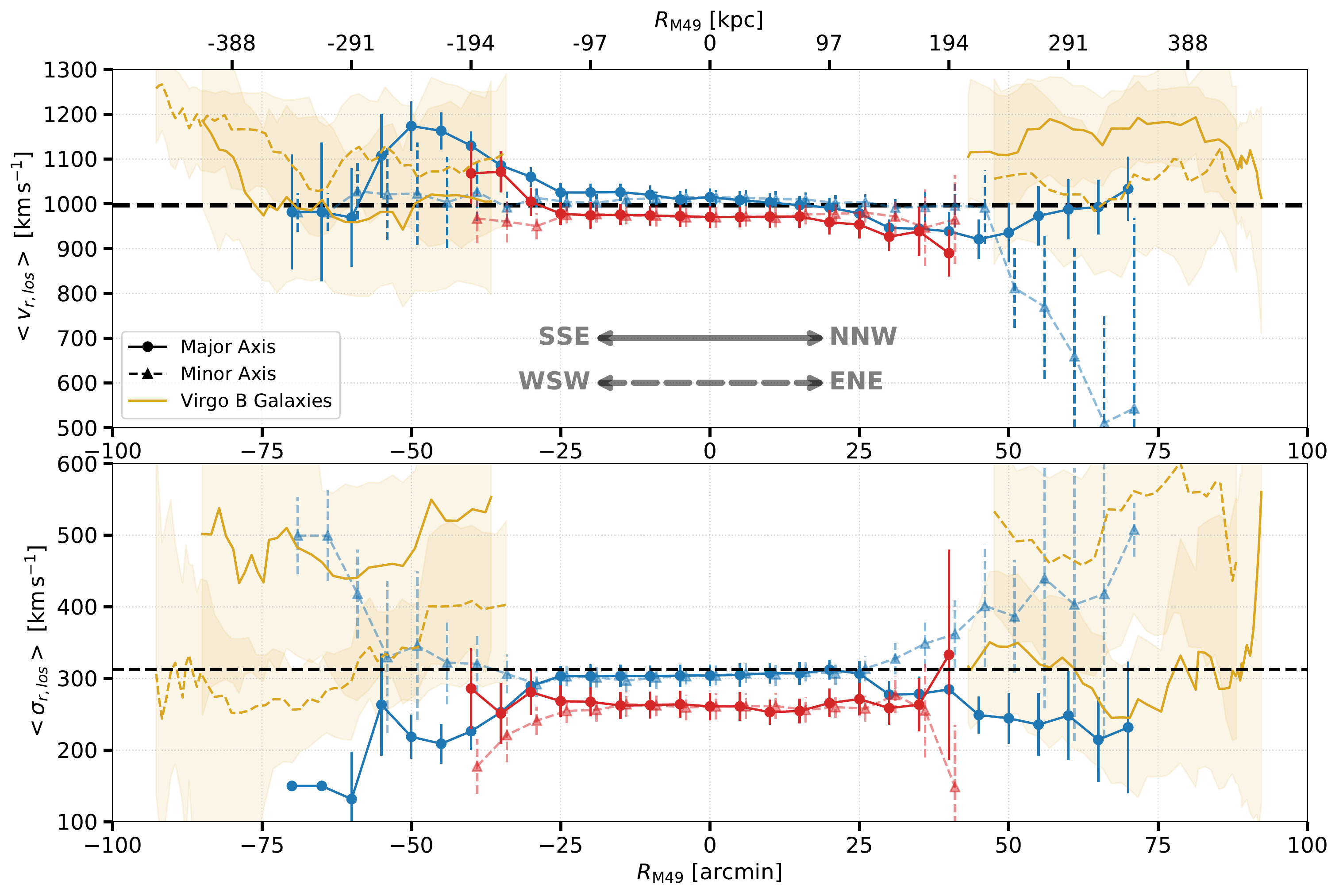}
\caption{One-dimensional \vrp\ (upper panel) and \vdrp\ (lower panel) profiles \citep[e.g.][]{lon18} sampled along the major (solid/circles) and minor (dashed/triangles) M49 photometric axes. Values for \rgc\ increase from SSE-NNW along the major axis, and from WSW-ENE along the minor axis. Blue and red GC profiles are colored accordingly, while gold relations/shading shows the same kinematic profiles derived in Fig.\,\ref{fig:vr_profile} but decomposed along each axis. Horizontal dashed black relations represent the $v_{r,los}=997\,\kps$ and $\sigma_{r,los}=303\,\kps$ of M49 itself. The decomposed axial profiles show hints of major axis rotation inside of $\sim50\arcmin$, with divergent large-scale \vdrp\ gradients that indicate a merging of GC kinematics with the surrounding Virgo B ICM.}
\label{fig:axis_losvd}
\end{figure*}

We further investigate the LOSVD in Fig.\,\ref{fig:axis_losvd}, where---following similar previous strategies \citep[e.g.,][]{lon18}---we decompose the \vrp\ (top panel) and \vdrp\ (bottom) profiles along the major (dots/solid lines; $\Phi=-21^\circ$ East of North) and minor (triangles/dashed lines; $\Phi=69^\circ$) axes. \rgc\ increases along the SSE-NNW and WSW-ENE directions, and we build the 1D profiles by sampling along each axis using a 30\arcmin-wide strip (i.e., selecting GCs within $\pm15\arcmin$ of each axis). The blue and red subsamples are colored accordingly, with the red sample truncated at $40\arcmin$ ($\sim200$\,kpc) owing to a dearth of red GCs with measured $v_{r,los}$ beyond this radius. For all samples, \vrp\ are consistent with the systemic M49 $v_{r,los}=997\,\kps$ in the inner $\sim25\arcmin$ ($\sim100$\,kpc), but each show a level of asymmetry at larger radii.

Looking at the major axis \vrp\ profile, a mild rotational signature is evident in both blue and red samples, with a peak-to-valley $\Delta v_{r,los}\approx200\,\kps$. A more detailed analysis of potential rotation at these radii will appear in a future work alongside those of other giant galaxies in the region. Along the major axis, both red and blue \vrp\ profiles behave similarly to the extent of the sampled red GCs, but beyond $\sim40\arcmin$ that of the blue GCs becomes asymmetric with the SSE sample showing a continued increase out to $\sim50\arcmin$ before plunging by $\sim200\,\kps$ beyond, albeit with large uncertainties due to sparse GC sampling. Meanwhile, on the other side of M49, the blue GC \vrp\ profile shows a steady positive gradient back to the systemic $v_{r,los}$ of M49.

Switching attention to the \vrp\ profile along M49's minor photometric axis, we again report very similar behavior between red and blue GCs to the extent of the red sample. We again see \vrp\ consistent with M49's systemic $v_{r,los}$ out to $\sim25\arcmin$, but beyond this there is a bifurcation at negative \rgc\ with those toward the WSW showing lower \vrp\ by $\ga100\,\kps$. However, red GCs sampled toward the ENE closely follow the major axis profile out to the limit of the sample such that the overall minor axis profile is generally flat in the sampled region. For the minor axis blue GCs, the profile is flat out to $\pm50\arcmin$, and are in excellent agreement with the $v_{r,los}$ of M49. Beyond $\sim50\arcmin$, the profile plummets in the NNW quadrant, although the sparesness of GCs and corresponding increase in \vrp\ uncertainties preclude firm conclusions to be drawn here.

The bottom panel of Fig.\,\ref{fig:axis_losvd} shows the same as the top, but for the \vdrp\ profiles. In all cases, the profiles exhibit very flat behavior in the inner $\sim25\arcmin$, with the blue GCs closely matching M49's $\sigma_{0}=303\,\kps$ (dashed black relation), while the red GCs show a similarly flat profile, but with kinematics cooler by $\sim50\,\kps$. Beyond $\sim25\arcmin$ the profiles begin to diverge. Similar to the \vrp\ profiles, the red GCs hint at a bifurcation, with those sampled along the major axis showing a mildly positive gradient in contrast with those sampled along the minor axis; however, we hesitate to make any strong inferences given the dearth of data points and associated uncertainties.

Like the red GCs beyond $\sim25\arcmin$, the blue sample shows divergent behavior between the two axes; however, both axes show large scale gradients that are also reflected in the PNe kinematics reported by \cite{pul18}. Here we see an overall symmetric cooling of the \vdrp\ profile along the major axis indicated by negative gradients to the extent of the sampled region. In contrast, we see that the positive gradient shown in Fig.\,\ref{fig:vr_profile} is largely due to the GCs that align with M49's minor photometric axis. Here the \vdrp\ profile rises symmetrically, with a peak-to-valley heating by $\sim200\,\kps$. As in Fig.\,\ref{fig:vr_profile}, we show the dynamics of the Virgo B subcluster galaxies as a rolling average, but decomposed along the overall direction of each axis. While there is a clear asymmetry in the Virgo B galaxy dynamics (likely due to the overall unrelaxed state of Virgo \citep[e.g.,][]{bin85, bin87, bin93}), the GC \vdrp\ profile rises to the $\sim400-500\,\kps$ level consistent with the overall signal shown by the galaxies. If we consider GCs sampled along the minor axis beyond $50\,\arcmin$, we find an overall $\sigma_{ICM}=425\pm60\,\kps$ that compares well with the ICM PNe sampled by \cite{har18}, who estimate $\sigma_{ICL}=397^{+38}_{-36}\,\kps$.

\begin{figure*}
\centering
\includegraphics[width=\linewidth]{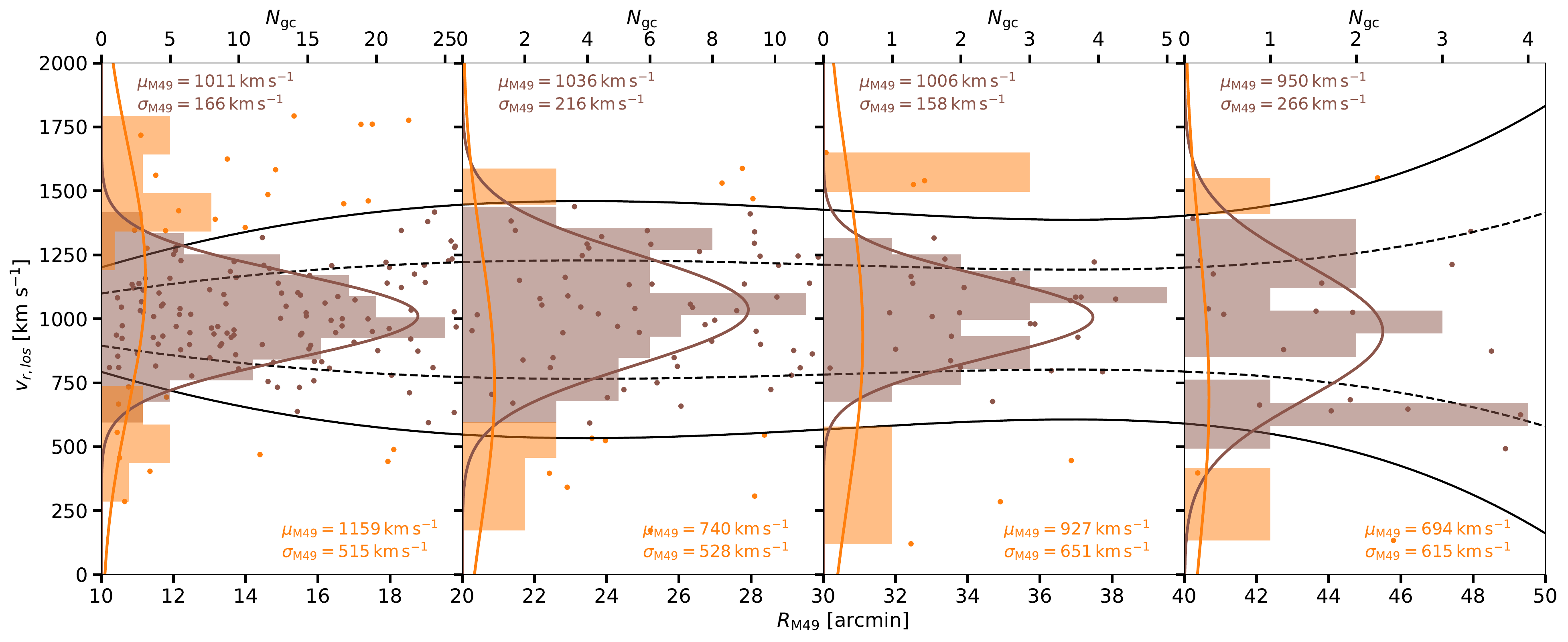}
\caption{Projected phase space in $10\arcmin < $\rgc$ < 50\arcmin$ ($50\la$\rgc$/{\rm kpc}\la 150$) where the ICM has increasing influence on GC dynamics. Brown and orange data indicate the distributions of GCs more likely to be dynamically influenced by M49 and the ICM, respectively, with Gaussian distribution parameters indicated. Dashed and solid black curves indicate robust $1\sigma$ and $2\sigma$ estimates used to classify the two populations \citep[e.g.,][]{lon15b}.}
\label{fig:robust_sig}
\end{figure*} 

To further investigate the transition from GCs dominated by M49 to those increasingly influenced by the ICM, we pay particular attention to the region  $10\arcmin < $\rgc$ < 50\arcmin$ ($50\la$\rgc$/{\rm kpc}\la 150$). Fig.\,\ref{fig:robust_sig} shows the projected phase-space for this region sampled in four equal 10\arcmin-wide bins. As an attempt to isolate the two populations, we adopt a ``robust sigma'' approach \citep[][]{mcn10,lon15b}. Briefly, in each bin we calculate the dispersion of the total GC sample, and mask out $2\sigma$ outliers with respect to M49's systemic $v_{r,sys}=997\,\kps$. From the survivors, we recalculate $\sigma$ and iterate using the Monte Carlo procedure outlined in \cite{mcn10} to rescale it to that of a complete Gaussian distribution. We apply this procedure in each of the four bins, and fit a fourth-order polynomial to the robust $\sigma$ estimates to correct for radial gradients. The dashed and solid black relations on Fig.\,\ref{fig:robust_sig} show the $1\sigma$ and $2\sigma$ thresholds from the polynomial fit, respectively, where we use the latter to identify individual GCs as either belonging to the population more likely to be under the influence of M49 itself (brown points/histogram) or that of the ICM (orange).

Continuing to follow the approach of \cite{lon15b}, we also show on Fig.\,\ref{fig:robust_sig} representative Gaussians for each population, with the observed $\mu$ and $\sigma$ listed in each bin/panel. We find 49 ICM GCs identified in this range, and integrating across each radial bin suggests that 8 more ICM GCs are indistinguishable from the main M49 sample, which implies 57 GCs in this transition range associated with the Virgo B subcluster ICM. If we consider the 30 GCs with projected \rgc\ beyond $50\arcmin$ to all be part of the ICM population (as suggested by Figs.\,\ref{fig:vr_profile}-\ref{fig:axis_losvd}), this brings our total number of implied ICM GCs to 87.

%%%%%%%%%%%%%%%%%%%%%%%%%%%%%%%%%%%%%%%%%%%%%%%%%%%%%%%%%%%%%%%%%
%%%%%%%%%%%%%%%%%%%%%%%%%%%%%%%%%%%%%%%%%%%%%%%%%%%%%%%%%%%%%%%%%
%%%%%%%%%%%%%%%%%%%%%%%%%%%%%%%%%%%%%%%%%%%%%%%%%%%%%%%%%%%%%%%%%

\subsection{GCs Associated with Shells}
\begin{figure*}
\centering
\includegraphics[width=\linewidth]{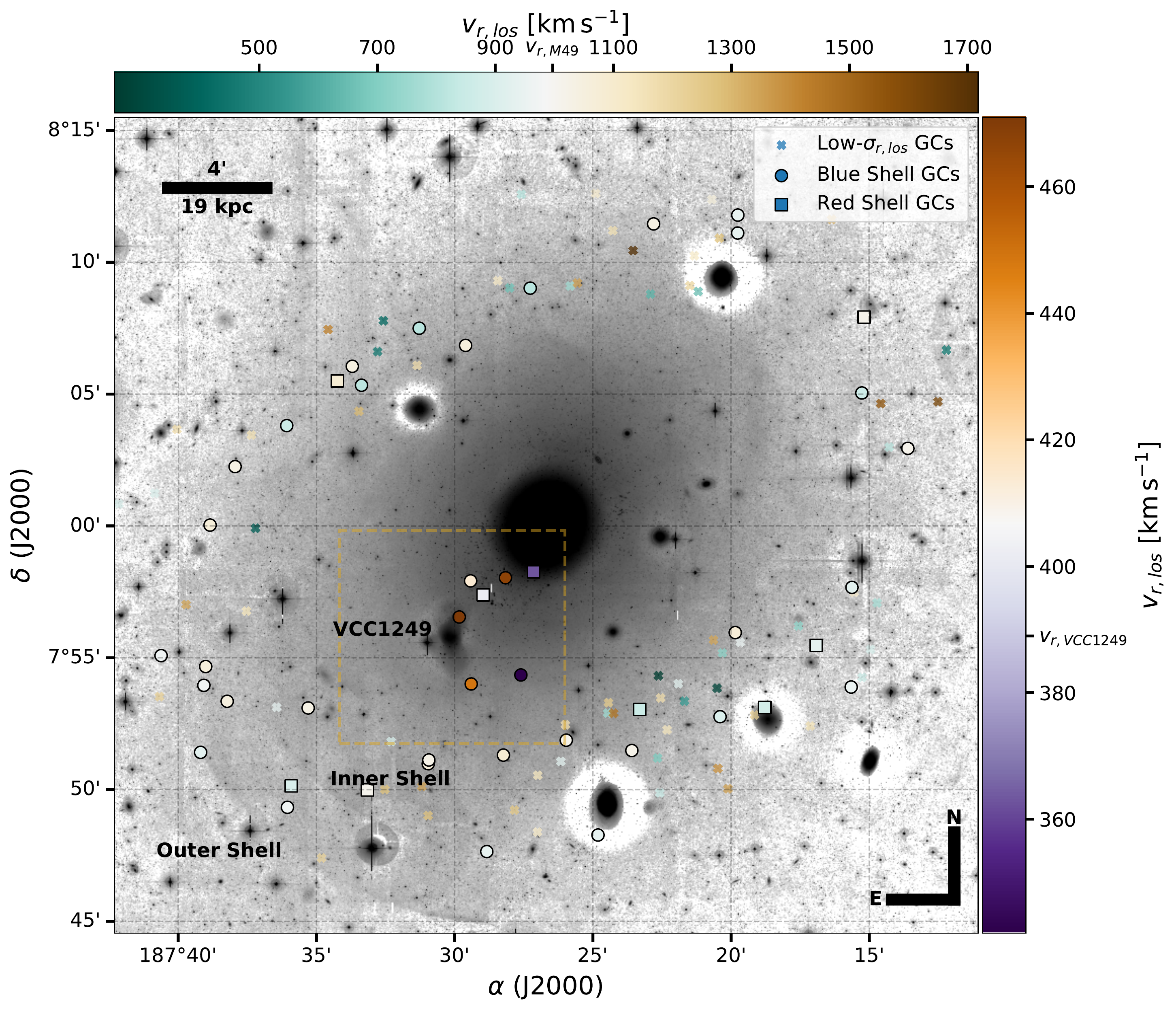}
\caption{A $g'$-band image from the NGVS, centered on M49 and unsharp masked to highlight LSB features, including VCC\,1249 and the two visible stellar shells to the Southeast. GCs in the elliptical annulus covering the \rgc\ range encompassing the kinematic ``dip'' anomaly in Fig.\,\ref{fig:vr_profile} are shown, with $v_{r,los}$ indicated by their colors as mapped to the upper color bar. For reference, we indicate by the yellow dashed-line box the zoom-in region centered on VCC\,1249 and displayed in Fig.\,\ref{fig:contour}, inside of which are GCs potentially associated with the dwarf galaxy with colors parametrizing $v_{r,los}$ as mapped to the right-hand color bar.  In each case, blue and red GCs are denoted by circular and square symbols, respectively. As indicated by the two colorbars, GCs inside the dashed box show $v_{r,los}$ are more representative of that shown by the nearby dwarf, while those outside are consistent with M49 and/or the visible shells to the SE.}
\label{fig:unsharp}
\end{figure*} 

The next feature in Fig.\,\ref{fig:vr_profile} is the rapid drop in the \vdrp\ profile shown by the cluster of points in $9\arcmin\!\la$\rgc$\!\la\!15\arcmin$, which exhibit the kinematically coolest signature of the entire sample. To investigate possible sources that might give rise to this feature, we limit GCs to those within this annulus and consider their distribution in $\Phi_{\rm M49}$ to search for evidence of clustering caused by a background galaxy projected near to M49. A K-S test provides no evidence to reject the notion that they are drawn from the same distribution as the overall sample, which would not be the case if these GCs were localized to a single host.

Fig.\,\ref{fig:unsharp} shows two isolated subsamples---one corresponding to the nearby dwarf galaxy VCC\,1249 (see \S\,\ref{sec:vcc1249}), and the other to the current region of interest---plotted over a $g'$-band image of M49. The image has been unsharp-masked to highlight sharp LSB features with a particular focus on two shells to the Southeast of M49 \citep{jan10}. All symbols outside of the dashed box (denoting the region shown in Fig.\,\ref{fig:contour}, see \S\,\ref{sec:vcc1249}) represent GCs that fall within the $9\arcmin\!\la$\rgc$\!\la\!15\arcmin$ annulus that straddles the ``dip'' in the \vdrp-profile, with color parameterizing $v_{r,los}$ as mapped to the upper color bar. Furthermore, we identify those likely to be associated with the stellar shells (see below) by the circles (blue GCs) and squares (red GCs).

We are interested in whether a portion of the GCs comprising the cold kinematic signature might be associated with the two stellar shells, and as an initial approach we isolate those that project on top of the visible shell arcs. Isolating those GCs in the range $90^\circ\leq \Phi_{\rm M49}\leq180^\circ$ results in 27 GCs (21 blue and six red) that collectively show \vdrp$=151\pm27\,\kps$. One outlying GC ($v_{r,los}\la400\,\kps$) is unlikely to be associated with the shells and its omission results in a somewhat lower \vdrp$=141\pm18\,\kps$. These GCs collectively make up the coolest kinematic feature seen in the data, and given their projection alongside the two visible shells makes a physical association plausible.

An expected signature of a GC accretion event is a chevron-like structure in the projected phase-space distribution \citep[e.g.,][]{rom12}. A robust search and statistical analysis of such a feature requires detailed simulations of M49's GC system, which is beyond the scope of this observational work. Despite this, we perform a search for the signature using a friends-of-friends group-finding approach \citep[e.g.,][]{per03, rud09}, as successfully implemented for the M87 system \citep[][]{rom12}. Briefly, this algorithm relies on a parametric distance $\Delta s$ defined by the parameters $w_R$ and $w_v$, weighted by \rgc\ and $v_{r,los}$, respectively. A third parameter is the linking length $\Lambda$ so that groups of objects in the projected phase-space can be identified by collections of objects that show $|\Delta s| < \Lambda$.

The group-finding results are sensitive to the adopted $w_r$, $w_v$, and $\Lambda$ parameters, which need to be chosen to highlight real features in the data while minimizing spurious groupings. This fine-tuning requires detailed modelling and simulations that are beyond the scope of this work, and we instead lean on \cite{rom12}, who tuned these parameters using simulations for the M87 system finding $w_r=0.035$\,dex, $w_v=50\,\kps$, and $\Lambda=0.07$. Given the similarities between M49 and M87 in terms of environment (i.e., dominant gEs of their respective Virgo subclusters) and spatial scales of interest (e.g., $50\la R/{\rm kpc}\la100$), we adopt these parameters in our investigation.

\begin{figure*}
\centering
\includegraphics[width=\linewidth]{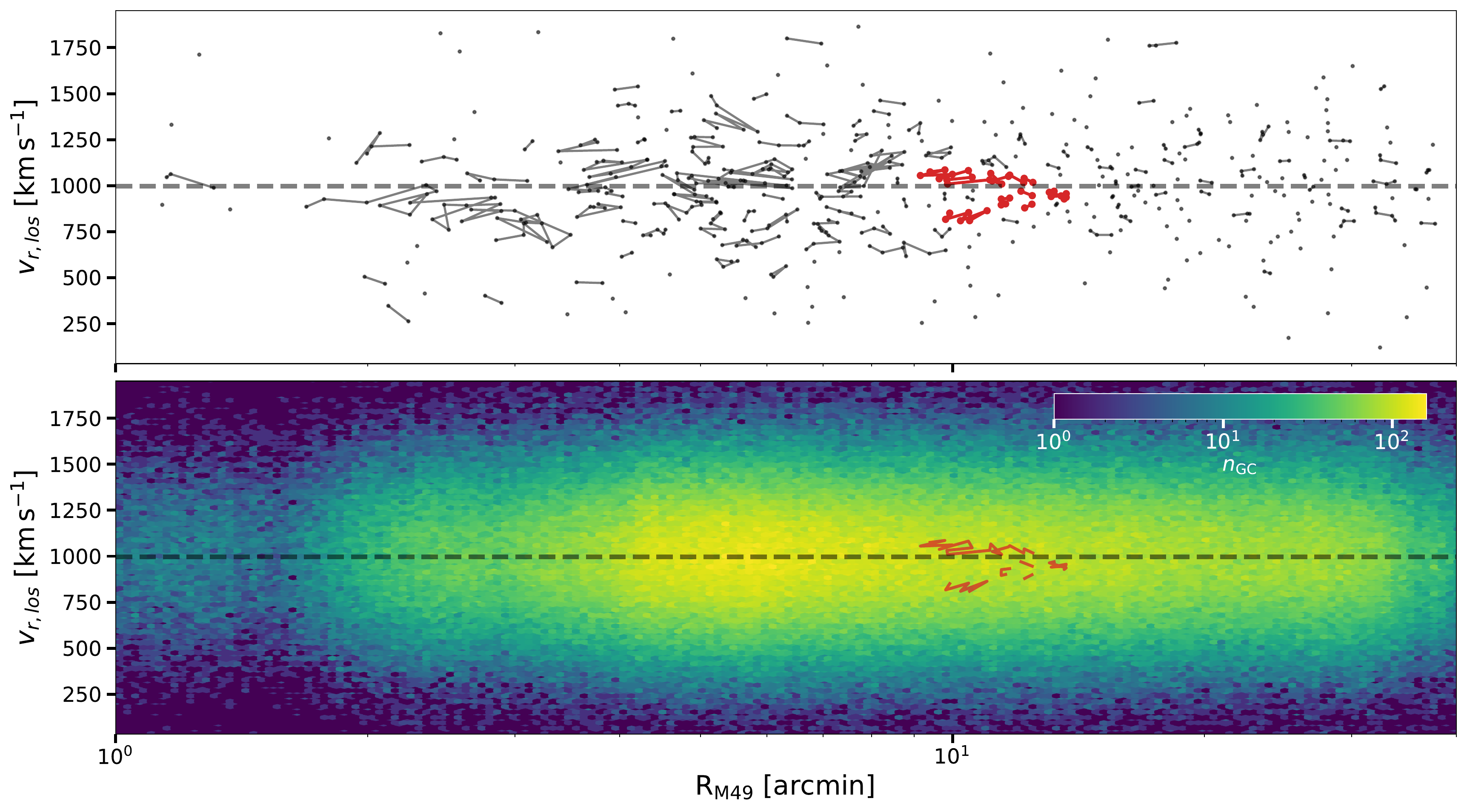}
\caption{Results of our group-finding algorithm and simulated phase space distribution. The upper panel shows the group-finding output assuming $w_R$, $w_v$, and $\lambda$ taken from the optimum parameters determined by \cite{rom12} for the M87 GC-shell system. Individual datapoints are indicated by dots, where we highlight a handful of groups that make up a chevron-like feature. In the bottom panel, we show the results of 1000 realizations of phase-space populated by simulated GCs. Color denotes the total number of GCs populating each cell in the diagram, and we note the lack of any obvious substructure in the region where the stellar shells are visible.}
\label{fig:groupfind}
\end{figure*} 

The upper panel of Fig.\,\ref{fig:groupfind} shows the outcome of the group-finding using our adopted $w_r$, $w_v$, and $\lambda$. The top panel shows the projected phase-space, where dots show individual GCs as linked by the algorithm output. We find a tentative detection of a handful of groups (highlighted red) that comprise a chevron-like feature with edges corresponding to the coolest kinematic feature discussed above, and terminating near the systemic velocity of M49 (horizontal dashed line). A formal estimate of the statistical significance of this feature requires detailed simulation work, but we show a qualitative assessment in the bottom panel of Fig.\,\ref{fig:groupfind}. Here we simulate the GC population by drawing from the observed spatial distribution of GCs and a $N(\mu_{obs}, \sigma_{obs})$ distribution in $v_{r,los}$ where $\mu_{obs}$ and $\sigma_{obs}$ are the observed mean and dispersion of our sample $v_{r,los}$. The colormap in the bottom panel of Fig.\,\ref{fig:groupfind} shows the number density of GCs that results from 1000 Monte Carlo realizations, with the groups making up the tentative chevron feature overplotted in red. While the observed shape mimics the overall shape of the phase-space envelope, if these particular groupings were common, they would form a recognizable structure in this region of phase-space. We conclude that, while far from a ``smoking gun'' detection of a clear kinematic fingerprint, this feature is at least consistent with these GCs being associated with a shell arising from a recent merger with M49 and/or the ongoing interaction between it and VCC\,1249.

We find a total of 38 GCs (seven of which are red) to make up this kinematic feature, which are a subset of the overall sample falling in the $9\arcmin\!\la$\rgc$\!\la\!15\arcmin$ annulus. We differentiate them by the squares/circles mapped to the upper color bar in Fig.\,\ref{fig:unsharp}. When we consider only this sample, we find the combined red/blue GCs to show a very cold \vdrp$=84\pm8\,\kps$, while those that project directly on the visible shells arcs (i.e. $90^\circ<\Phi_{\rm M49}<180^\circ$) exhibit a remarkably low \vdrp$=61\pm10\,\kps$. Given that these GCs make up a tentative kinematic fingerprint of a recent accretion event, and that they represent the coldest kinematic signature driving the dip in the \vdrp-profile seen in Fig.\,\ref{fig:vr_profile}, we consider it plausible or even likely that at least a significant fraction of these 38 GCs are caught in the act of being accreted onto M49's outer halo.

%%%%%%%%%%%%%%%%%%%%%%%%%%%%%%%%%%%%%%%%%%%%%%%%%%%%%%%%%%%%%%%%%
%%%%%%%%%%%%%%%%%%%%%%%%%%%%%%%%%%%%%%%%%%%%%%%%%%%%%%%%%%%%%%%%%
%%%%%%%%%%%%%%%%%%%%%%%%%%%%%%%%%%%%%%%%%%%%%%%%%%%%%%%%%%%%%%%%%

\subsection{GCs Associated with VCC\,1249 \label{sec:vcc1249}}
\begin{figure}
\centering
\includegraphics[width=\linewidth]{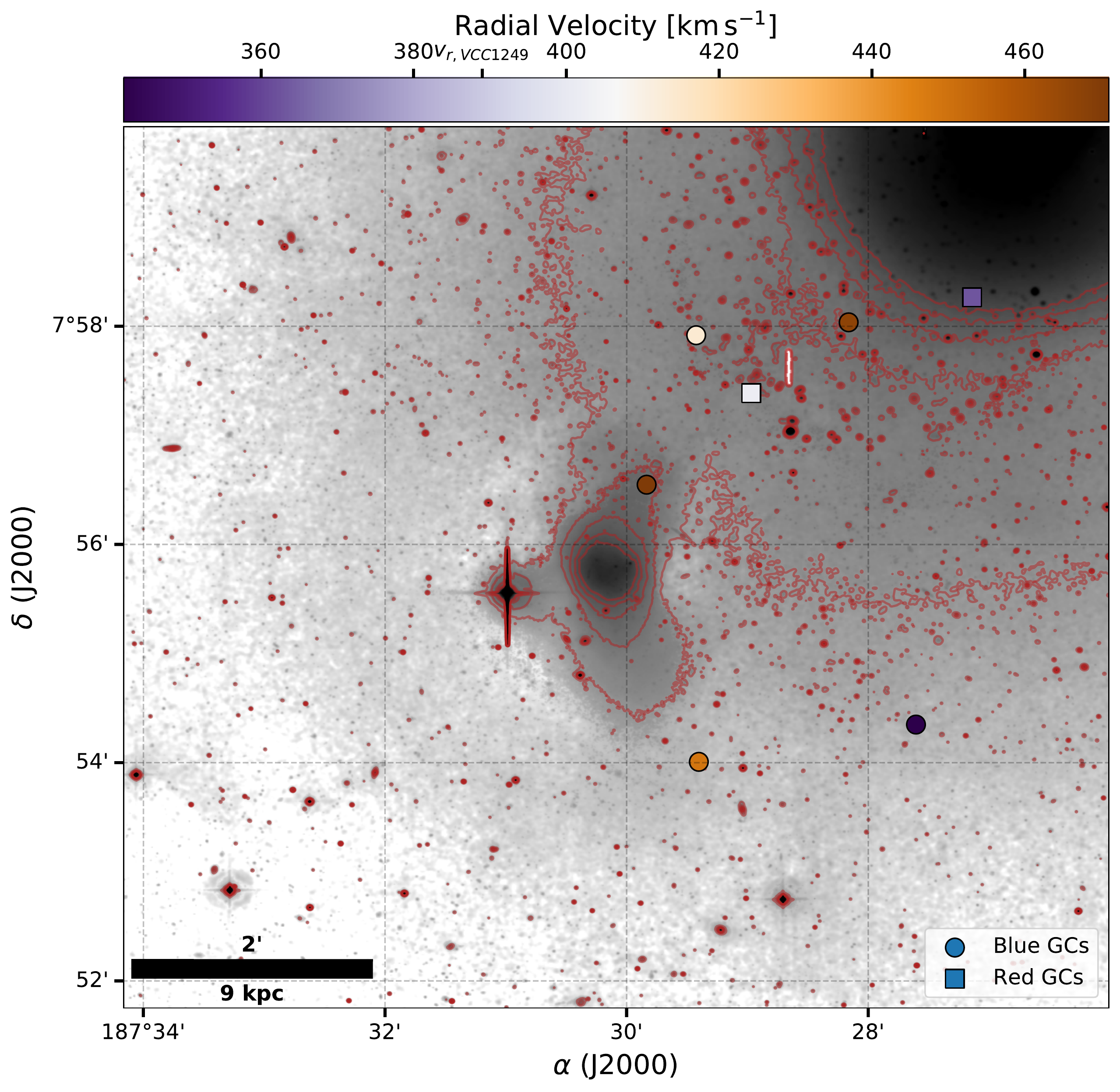}
\caption{A cutout image corresponding to the yellow dashed box displayed on Fig.\,\ref{fig:unsharp}, centered on VCC\,1249 and with isophotal contours oveprlotted in red. Circles and squares indicate five blue and two red GCs, respectively, with $v_{r,los}$ consistent with VCC\,1249 itself.}
\label{fig:contour}
\end{figure} 

The ongoing interaction \citep[e.g.][]{arr12} between M49 and the nearby (5.4\arcmin; 26\,kpc) dwarf galaxy VCC\,1249, presents an excellent opportunity to identify GCs in the act of being accreted onto the giant halo. \cite{har18} report a kinematic feature marked by a cluster of PNe likely to be associated with the dwarf that are spatially coincident, share a systemic $v_{r,los}$ indistinguishable from VCC\,1249 itself, and show a collective $\sigma_{\rm PNe, VCC1249}=40^{+28}_{-19}\,\kps$.

While \cite{har18} were able to isolate a kinematically cooler subset of PNe corresponding to VCC\,1249's $\sigma_{\rm VCC\,1249}\approx40\,\kps$, we find no clear group of GCs that correspond directly in phase-space with the dwarf. However, recognizing that GCs may be in the process of being stripped from VCC\,1249 (i.e.\ may not exactly coincide spatially), we search for candidates in a similar fashion to that described in \S\,\ref{sec:gc_sample}. Specifically, we isolate GCs lying within five \reff\ of the dwarf \citep[$r_{\rm eff, VCC\,1249}=48\arcsec$; 3.65\,kpc;][]{fer20} and with $|v_{r,los} - v_{r,{\rm VCC\,1249}}| < 3\sigma_{0,{\rm VCC\,1249}}$. We find seven GCs meeting these criteria, including two red and five blue, which are indicated on Fig.\,\ref{fig:unsharp} by circles (blue GCs) and squares (red) nearest the dwarf galaxy. Color parameterizes GC $v_{r,los}$ as mapped to the right-hand color gradient. These GCs are not uniformly distributed around VCC\,1249, showing a stream-like distribution toward M49, and no candidates found Eastward of the dwarf.

Fig.\,\ref{fig:contour} shows a zoom-in corresponding to the dashed box shown on Fig.\,\ref{fig:unsharp}, centered on VCC\,1249, with a field of view of ten $r_{\rm eff, VCC\,1249}$ per side. Overplotted are the seven VCC\,1249 GC candidates, as well as isophotal contours (red curves). Noteworthy in Fig.\,\ref{fig:contour} is the clear banana-like shape of the VCC\,1249 isophotes, closely approximating the overall boomerang shape of the GC distribution. A stream-like feature can be expected from GCs being stripped in such an interaction and based on the similar morphologies shown by the GCs and VCC\,1249 isophotes, we posit that we are witnessing these GCs actively being accreted into M49's halo.

In the crowded environment surrounding M49, there may be a non-negligible chance that we select GCs that happen to spatially coincide with VCC\,1249, while also sharing similar kinematics. To quantify this chance, we carry out 10\,000 Monte Carlo realizations mimicking the selection of the VCC\,1249 GC candidates and consider the absolute number of GCs selected in each realization, as well as their overall spatial distributions. For each realization, we generate a random set of $(\alpha, \delta)$ coordinates within five $r_{\rm eff,M49}$ of M49 itself (to only sample regions of similar GC density as VCC\,1249) and (as done for the VCC\,1249 GC candidate selection) isolate those within five $r_{\rm eff, VCC1249}$ of that point with $|v_r - v_{r,{\rm VCC1249}}| \leq 3\sigma_{r,{\rm VCC1249}}$. We further consider only those realizations that result in $N_{gc}=7\pm2$ GCs, resulting in 537 (5.37 percent) samples that meet this criteria. The elongated distribution of the VCC\,1249 GC candidates provides another constraint, and to account for this we find the best fitting ellipse whose edge follows the GC distribution and consider its ellipticity, $e_{\rm samp}$, in comparison to that describing a fit to the VCC\,1249 candidates, $e_{\rm VCC1249}$. Cases where $e_{\rm samp} > e_{\rm VCC1249}$ imply more elongated distributions and taking this into account alongside $N_{gc}$, results in 82 (0.82 percent) samples with similarly elongated shapes. We conclude that at least some of these GCs are very likely to be members of the VCC\,1249 GC system currently being accreted by M49 itself.

%%%%%%%%%%%%%%%%%%%%%%%%%%%%%%%%%%%%%%%%%%%%%%%%%%%%%%%%%%%%%%%%%
%%%%%%%%%%%%%%%%%%%%%%%%%%%%%%%%%%%%%%%%%%%%%%%%%%%%%%%%%%%%%%%%%
%%%%%%%%%%%%%%%%%%%%%%%%%%%%%%%%%%%%%%%%%%%%%%%%%%%%%%%%%%%%%%%%%

\section{Summary and Conclusions} \label{sec:conc}
We have presented the first results from a spectroscopic survey of GCs around the brightest galaxy in the Virgo galaxy cluster, M49. We use spatial and velocity information to clean the sample of GCs belonging to the myriad galaxies in the field, resulting in a sample consisting of GCs associated with M49 itself, or which have no clear association with any galaxy in the Virgo B subcluster. We further classified GCs as either blue (metal-poor) or red (metal-rich), and analyzed their kinematic profiles alongside the overall sample.

Beyond $\sim100$\,kpc of M49, the overall mean \vrp\ profile declines and the kinematics become hotter, as evidenced by a positive \vdrp\ gradient. We further decompose the kinematic profiles along the major and minor photometric axes of M49, and find both symmetric and asymmetric M49-centric profiles depending on both GC color and axis subsamples. We find that the hotter kinematics in the extreme halo of M49 is dominated by blue GCs sampled along M49's minor axis, with a positive \vdrp\ gradient that smoothly merges with the kinematics of the Virgo B subcluster. We argue that this implies a transition from GCs dominated by the potential well of M49 to that defined by the Virgo B ICM. We apply a radially dependent robust sigma approach to further investigate this transition region, and estimate the number of GCs in our sample that are most likely to be kinematically ruled by the larger subcluster dynamics.

A notable dip in the overall \vdrp\ profile reveals a population of GCs with particularly cold kinematics, a subsample of which are spatially co-located with the two stellar shells in projection. From these GCs, we find a hint of a kinematic feature in projected phase-space, and identify the most likely GCs to be genuinely associated with the visible shells. The blue colors of these GCs, combined with their coherent kinematics and spatial alignment suggests that at least some of these GCs have been caught in the act of being accreted onto M49's halo.

Finally, we report the spatial and kinematic alignment of a group of seven GCs that coincide with the nearby dwarf galaxy VCC\,1249, which is currently interacting with M49 itself. Isolating GCs with $v_{r,los}$ consistent with the dwarf, and projecting nearby reveals a stream stretching toward M49, with a distribution that mimics the isophotal contours of VCC\,1249. We attempt to quantify the statistical significance of the kinematic/spatial alignment with the dwarf, finding them to be significant at the $3\sigma$ level, suggesting that msot of these GCs are currently being stripped from VCC\,1249 as it passes through M49's halo.

Taken together, these data appear to straddle the transition from the GC system intrinsic to the dominant galaxy of the Virgo B subcluster to those ruled by Virgo ICM dynamics. Additionally, we seem to have come across the rare opportunity to study a giant galaxy caught in the act of accreting GCs into its outer halo. Further exploitation of these data will shed valuable light on the process of halo mass assembly of a giant galaxy in a cluster environment.

\acknowledgments
We extend our gratitude to the anonymous referee, whose detailed comments  served to greatly improve the original work. We also thank J.\,Hartke, who kindly provided data allowing a direct comparison to their previous work. M.A.T., P.C., L.F., and J.R.\ are supported by the National Research Council of Canada (NRC). I.C.'s research is supported by the Smithsonian Astrophysical Observatory Telescope Data Center, the Russian Science Foundation grant 17-72-20119 and the Program of the Interdisciplinary Scientific and Educational School of MV.\ Lomonosov Moscow State University ``Fundamental and Applied Space Research''. C.L. acknowledges support from the NSFC, Grant No. 11673017, 11833005, 11933003. This research uses data obtained through the Telescope Access Program (TAP), which has been funded by the National Astronomical Observatories, Chinese Academy of Sciences, and the Special Fund for Astronomy from the Ministry of Finance. Observations reported here were obtained at the MMT Observatory, a joint facility of the University of Arizona and the Smithsonian Institution and on observations obtained with MegaPrime/MegaCam, a joint project of the Canada-France-Hawaii Telescope (CFHT) and CEA/DAPNIA, at the CFHT which is operated by the NRC, the Institut National des Sciences de l’Univers of the Centre National de la Recherche Scientifique of France, and the University of Hawaii. We thank Perry Berlind, Erin Martin, and the MMT staff for their professional support of these observations. This research has made use of NASA's Astrophysics Data System Bibliographic Services, and NASA's Extragalactic Data System.

\vspace{5mm}
\facilities{MMT(Hectospec), CFHT}

\software{astropy \citep{ast13},
	matplotlib \citep{hun07},
	numpy \citep{van11}
	scikit-learn \citep{ped11},
	scipy \citep{vir20},
	topcat \citep{tay05}
}

\end{document}